\documentclass[twocolumn,nofootinbib,longbibliography,prx,superscriptaddress]{revtex4-1}

\usepackage{feynmp}
\usepackage{amsmath}
\usepackage{graphicx}
\usepackage{multirow}
\usepackage{latexsym}
\usepackage{textcomp}
\usepackage{verbatim}
\usepackage{color}
\usepackage{bm}
\usepackage{subfigure}
\usepackage{everysel}
\usepackage{keyval}
\usepackage{ragged2e}
\usepackage{dsfont}
\usepackage{amssymb}
\usepackage{enumitem}
\usepackage{pstricks}
\usepackage{mathtools}
\usepackage{braket}
\usepackage{slashed} 
\usepackage[colorlinks=true]{hyperref} 
\usepackage{placeins} 
\usepackage{cleveref}

\usepackage{xr}

\usepackage[vcentermath]{youngtab}
\usepackage{ytableau}
\usepackage{float}
\usepackage{tabularx}
\usepackage{adjustbox}

\usepackage{MnSymbol}

\usepackage{graphicx}
\usepackage{xcolor}
\usepackage{amsmath}
\usepackage{amsfonts}
\usepackage{bbm}

\usepackage{ulem}

\newcommand{\osum}{{%
    \setbox0\hbox{\circ}%
    \rlap{\hbox to \wd0{\hss\sum\hss}}\box0
}}

\newcommand{\tr}{{\textrm{tr}}}

\newcommand{\cO}{{\mathcal{O}}}

\hypersetup{
	colorlinks=true,
	linkcolor=blue,
	filecolor=magenta,      
	urlcolor=blue,
	citecolor=blue
}

\begin{document}

\title[Article Title]{L-entropy: A new genuine multipartite entanglement measure}


\author{Jaydeep Kumar Basak}
\thanks{Electronic Address: jkb.hep@gmail.com}
\affiliation{Department of Physics, Gwangju Institute of Science and Technology,
123 Cheomdan-gwagiro, Gwangju 61005, Korea}

\author{Vinay Malvimat}
\thanks{Electronic Address: vinaymalvimat@khu.ac.kr}
\affiliation{Department of Physics, College of Science, Kyung Hee University, Seoul 02447, Republic of Korea}

\author{Junggi Yoon}
\thanks{Electronic Address: junggi.yoon@khu.ac.kr}
\affiliation{Department of Physics, College of Science, Kyung Hee University, Seoul 02447, Republic of Korea}

\begin{abstract}

We advance  ``Latent entropy" (L-entropy) as a novel measure to characterize genuine multipartite entanglement in pure states, applicable to quantum systems with both finite and infinite degrees of freedom. This measure, derived from an upper bound on reflected entropy, attains its maximum for three-party GHZ states and $n=4,5$-party $2$-uniform states. We establish that it satisfies all essential properties of a genuine multipartite entanglement measure, including being a pure-state entanglement monotone.  We further obtain an analogue of the Page curve by analyzing the behavior of L-entropy in multiboundary wormholes, emphasizing their connection to multipartite entanglement in random states. Specifically, for $n = 5$, we show that random states approximate $2$-uniform states, exhibiting maximal multipartite entanglement. Extending these ideas to finite temperatures, we introduce the Multipartite Thermal Pure Quantum (MTPQ) state, a generalization of the thermal pure quantum state to multipartite systems, and demonstrate that the entanglement structure in states of the multicopy SYK model exhibits finite-temperature $2$-uniform behavior.


\end{abstract}

\maketitle

\textit{Introduction.}---Multipartite entanglement holds immense significance for a variety of fascinating phenomena ranging from quantum computers to black hole information loss paradox. Most physically relevant systems naturally exhibit multipartite entanglement, making its characterization essential for understanding fundamental aspects of quantum information theory. Numerous studies have shown that multipartite entanglement can significantly enhance the efficiency of various quantum protocols, including quantum teleportation, quantum secret sharing, and even quantum batteries \cite{Ma:2023ecg,Horodecki:2024bgc,Choi:2022lge,Hillery:1998yq,Cleve:1999qg,Shi:2025gyu}. Measurement-based quantum computing (MBQC) is an entire paradigm of quantum computation that relies on multipartite entanglement as a fundamental resource\cite{briegel2009measurement,Wong:2022mnv}.  Furthermore, multipartite entanglement has also become crucial in understanding deeper aspects of black holes \cite{Gadde:2022cqi,Iizuka:2024pzm}. In this context, it becomes crucial to characterize the multipartite entanglement present in the quantum state.

An n-party pure state is said to possess genuine multipartite entanglement if it cannot be factorized across any bipartition. As shown in \cite{Dur:2000zz}, even in the simplest case of three qubits, there exist two inequivalent classes of states with genuine tripartite entanglement. Their classification was based on the equivalence of the states under local operations and classical communication with non-vanishing probability (SLOCC).  As the number of parties increases, the number of such inequivalent classes grows rapidly, making the characterization of multipartite entanglement a highly complex problem. Note that any measure based solely on the entanglement entropies of various subsystems is insufficient due to the existence of iso-spectral states belonging to different SLOCC classes. These states have the same entanglement entropy values for all the subsystems despite belonging to entirely distinct classes and hence they demand new measures for characterizing multipartite entanglement. 

To establish a consistent framework for quantifying genuine multipartite entanglement (GME), three fundamental axioms have been proposed that any valid measure must satisfy \cite{Plenio:2007zz,Ma:2023ecg}. A subsequent work introduced a weaker supplementary criterion which, while not essential, further refines the classification of proper GME measures from comparing the GHZ and $W$ states in three-party systems, based on their relative efficiency in quantum teleportation \cite{PhysRevLett.127.040403}. Several GME quantifiers have been developed \cite{Coffman:1999jd,Sabin:2008naa,PhysRevA.80.032330,Ma:2011yon,PhysRevA.81.012308,PhysRevLett.113.110501,PhysRevLett.127.040403,Xie:2023dzw}, though most are applicable only to systems with finite degrees of freedom or a limited number of parties. For example, the genuinely multipartite concurrence \cite{Ma:2011yon} and the generalized geometric measure \cite{PhysRevA.81.012308} apply to arbitrary $n$-party states but are confined to finite-dimensional Hilbert spaces. The concurrence fill, in contrast, was proposed as a genuine tripartite measure \cite{PhysRevLett.127.040403}, while its four-party analogue, the entropy fill, extends to infinite-dimensional systems via specific combinations of entanglement entropies \cite{Xie:2023dzw}. However, the latter remains restricted to four-party configurations, lacking a straightforward generalization to arbitrary $n$. Moreover, measures constructed solely from entanglement entropies fail to distinguish isospectral states whose $k$-party reduced density matrices possess identical eigenvalue spectra and thus yield the same entanglement entropies, despite belonging to distinct entanglement classes (see for example the discussion in \cite{Gadde:2022cqi}).

In this letter, we introduce a novel measure based on an upper bound for reflected entropy, called latent entropy (L-entropy), to address the challenge of characterizing genuine multipartite entanglement in pure states. Note that, unlike most existing measures, our measure is applicable not only to quantum systems with finite-dimensional  Hilbert spaces but also to systems with infinite degrees of freedom such as quantum field theories (QFTs), conformal field theories (CFTs), and holographic settings. 

We first establish that our measure satisfies all the standard criteria required in quantum information theory. For three qubits, it attains its maximum for the GHZ state, identifying it as a 1-uniform state. A $k$-uniform state is an $n$-party pure state whose all $k$-party reduced density matrices are maximally mixed. It is generally expected that $\lfloor n/2 \rfloor$-uniform states exhibit maximal multipartite entanglement \cite{Scott:2004vaq,PhysRevA.86.052335}, implying that the GHZ state carries the maximal GME among three-party systems. This trend persists for higher $n$ and larger Hilbert space dimensions, where the measure peaks for 2-uniform states. Hence, the generalized L-entropy for arbitrary $n$-party pure states is expected to maximize for $\lfloor n/2 \rfloor$-uniform configurations \cite{newlentropy}. Extending this notion to finite temperatures, we show that a multi-copy SYK model approximates such a finite-temperature 2-uniform state. We also evaluate our measure in a multiboundary wormhole setup and obtain an analogue of the Page curve for genuine multipartite entanglement.

\textit{GME and reflected entropy.}---In this section, we briefly review the properties that any GME measure is expected to satisfy. We then outline the definition and relevant properties of the reflected entropy as they pertain to this letter. As discussed in the Introduction, an n-partite state is said to exhibit genuine multipartite entanglement if it cannot be factorized across any bipartition. A GME measure ${\cal E}$ is expected to satisfy the following properties:
\begin{enumerate}
	\item ${\cal E}=0$ if and only if the state is separable
	\item ${\cal E}$ is invariant under local unitary (LU) operations.
	\item ${\cal E}$  obeys montonicity under local operations and classical communication (LOCC) on average.
\end{enumerate}
Any measure satisfying the above conditions is referred to as a GME measure. Additionally, if the measure ranks the GHZ state higher than the W state for three parties, it is termed a proper GME measure \cite{PhysRevLett.127.040403}.

Let us begin with a pure multipartite state $\ket{\psi}_{AB \overline{AB}}$. The bipartite state $\rho_{AB}$ can be obtained from such a state by tracing out $\overline{AB}$
\begin{align}
	\rho_{AB}=Tr_{\overline{AB}}(\ket{\psi}\bra{\psi})
\end{align}
In order to define the reflected entropy, one needs to purify the given mixed $\rho_{AB}$ state in a canonical way by doubling the Hilbert space, considering $A^*$ and $B^*$ as copies of $A$ and $B$, respectively. This canonically purified state is denoted as $\ket{\sqrt{\rho_{AB}}}_{ABA^*B^*}$, and the reflected entropy is defined as the von Neumann entropy of the subsystem $AA^*$ or $BB^*$ in that state \cite{Dutta:2019gen}
\begin{align}
	S_R(A:B)=S(AA^*)_{\ket{\mathrm{\small\sqrt{\rho_{AB}}}}}=S(BB^*)_{\ket{\sqrt{\rho_{AB}}}}
\end{align}
Using the subadditivity and strong subadditivity properties between the various subsystems $A$, $A^*$, $B$, and $B^*$, it can be straightforwardly demonstrated that the reflected entropy is bounded from above and below as follows:
\begin{align}\label{ref_bound}
	I(A:B) \leq S_R(A:B)\leq \textrm{Min}\{2S(A),2S(B)\}
\end{align}
where $I(A:B)=S(A)+S(B)-S(AB)$ is the mutual information between $A$ and $B$. The lower bound derived from the strong subadditivity is related to the Markov recovery process of the tripartite state $AA^*B$ from the bipartite state using only LOCC operations. Consequently, the difference between the reflected entropy and its lower bound was proposed as a measure of the tripartite entanglement in \cite{Akers:2019gcv,Hayden:2021gno}. This quantity, termed the Markov gap, has been extensively studied in recent years \cite{Zou:2020bly}.

\textit{A new measure.}---In this section, we introduce a novel measure, \textit{latent entropy} (L-entropy), to quantify genuine $n$-partite entanglement in a pure state for $n\leq 5$. Hereafter, we restrict our discussion to $n$-partite states and L-entropy with $n\leq5$, unless stated otherwise. Now, we propose the definition of $n$-partite L-entropy for a $n$-party pure state $|\psi\rangle_{A_1A_2\cdot\cdot\cdot A_n}$ as
\begin{equation}\label{multi_l}
	\ell_{A_1A_2\cdot\cdot\cdot A_n}=\left(\prod_{i<j}\ell_{A_iA_j}\right)^{\frac{2}{n(n-1)}}.
\end{equation}
Here, the bipartite L-entropy, $\ell_{A_i,A_j}$ is expressed as,
\begin{align}
	\ell_{A_i,A_j}&=\textrm{Min}\{2S(A_i),2S(A_j)\}-S_R(A_i:A_j)\label{bip_l_1}\\
	&=\textrm{Min}\{I(A_i:A_i^*),I(B_j:B_j^*)\}\label{bip_l_2}
\end{align}
with the indices, $i=1,2,\cdot\cdot\cdot,n$, $j=i+1,i+2,\cdot\cdot\cdot,n$ which consideres all possible bipartitions. Following the upper bound of reflected entropy in \cref{ref_bound}, the bipartite and multipartite L-entropy are semi-positive. For an arbitrary bipartition, the bound on the bipartite L-entropy can be obtained as $\ell_{A_iA_j}\leq \textrm{Min}\{ 2\log[d_{A_i}],2\log[d_{A_j}],\log[d_{\overline{A_iA_j}}] \}$ where $d_{A_i}, d_{A_j}, d_{\overline{A_iA_j}}$ are the dimensions of the Hilbert spaces corresponding to $A_i, A_j, \overline{A_iA_j}$ respectively. Considering a product state $|\psi\rangle_{A_1A_2\cdot\cdot\cdot A_n}=|\phi_1\rangle _{A_1}\otimes |\phi_2\rangle_{A_2} \otimes \cdot\cdot\cdot \otimes |\phi_n\rangle_{A_n}$, the entanglement entropy of each subsystem, as well as the reflected entropies between any two subsystems, are identically zero. Consequently, both bipartite and multipartite L-entropies vanish. In the case of biseparable states, at least one of the bipartite L-entropy becomes zero, leading to a vanishing multipartite L-entropy. This guarantees compliance with Condition (1). Note that we have only established one direction of the condition namely, that the L-entropy vanishes for all biseparable states. We conjecture that it remains nonzero for any GME state and, so far, have not encountered any counterexamples. Condition (2) can be checked by applying LU transformations on the subsystems, which modify the canonically purified state as  
\begin{align}
	\ket{\sqrt{\rho_{A_iA_j}'}}=U_{A_i} \otimes U_{A_j}\otimes U_{A_i^*} \otimes U_{A_j^*} \ket{\sqrt{\rho_{A_iA_j}}}.
\end{align}
Here $U_{A_i^*}$ and $U_{A_j^*}$ are copies of the unitaries $U_{A_i}$ and $U_{A_j}$ respectively. Following \cref{bip_l_2}, the LU invariance of the entanglement entropy for $A_i, A_j, A_i^*$ and $A_j^*$ indicates that the multipartite L-entropy is invariant under LU  satisfying condition (2). Finally, Condition (3) is obtained as each $\ell_{A_i,A_j}$ is shown to be entanglement monotone and the geometric mean of entanglement monotones also preserves monotonicity \cite{Li:2021hhj}. See Supplimentary Material \cite{suppl} for the proof.

Now we consider the simplest example of a multipartite state involving three qubits. Such a tripartite pure state belongs to one of two distinct classes depending on the behaviour under SLOCC: the W class and the GHZ class \cite{Dur:2000zz}. The W state $\left|W\right\rangle=(|001\rangle+|010  \rangle+|100\rangle)/\sqrt{3} $ and the GHZ state $\left|GHZ\right\rangle=(|000\rangle+|111  \rangle)/\sqrt{2} $ serve as the most interesting representative examples of the W and GHZ classes, respectively. Upon tracing out any one of the constituent parties, the W state retains maximum residual entanglement between the remaining two parties, whereas the GHZ state renders them separable. In this context, tripartite L-entropy indicates that the GHZ state manifests a considerably higher degree of tripartite entanglement with $ \ell_{ABC}(|GHZ\rangle) = 1 $ which is also the maximum permissible tripartite L-entropy, compared to the W state where $ \ell_{ABC}(|W\rangle) = 0.35 $ in $\log 2$ units.
Consequently, due to this property, the L-entropy further qualifies as a proper GME measure complementing other measures reported in the literature, such as the concurrence fill $ F_{123} $ \cite{PhysRevLett.127.040403} and the GME-concurrence $C_{GME}$ \cite{Ma:2011yon}, where the values are given as $ F_{123}(|W\rangle) = \frac{8}{9} = C_{GME}(|W\rangle) $ and $ F_{123}(|GHZ\rangle) = 1 = C_{GME}(|GHZ\rangle) $. Interestingly, L-entropy further suggests even a lower amount of genuine tripartite entanglement in the W state compared to the results in \cite{PhysRevLett.127.040403,Ma:2011yon}.

Although the definition for the multipartite L-entropy in \cref{multi_l} is given for pure states, it is possible to compute the genuine $n$-party entanglement present in a $n$-party mixed state utilizing the convex roof extension \cite{Ma:2023ecg,Ge:2022sqp} of L-entropy where the mixed state can be written as a weighted sum of pure states as $\rho_{A_1A_2\cdot\cdot\cdot A_n}=\sum_i p_i\left|\psi_i\right\rangle\left\langle\psi_i\right|$ for $\sum_i p_i=1$. Here, the $n$-party L-entropy is expressed as,
\begin{align}
	\operatorname{CR}_\ell(\rho):=\min _{M_\rho} \sum_i p_i \ell_{A_1A_2\cdot\cdot\cdot A_n}\left(\left|\psi_i\right\rangle\right), 
\end{align}
where $\min _{M_\rho}$ denotes the minimization over all possible pure state decompositions of the density matrix.

Although the definition of the L-entropy has been restricted up to five parties in the present letter, the construction can be generalized to define a GME measure for an arbitrary $n$-party pure state \cite{newlentropy}. See Supplementary Material  for an illustrative formulation of this generalization \cite{suppl}.

\begin{figure}[t]
	\centering
	\begin{adjustbox}{center}
		\includegraphics[width=.8\linewidth]{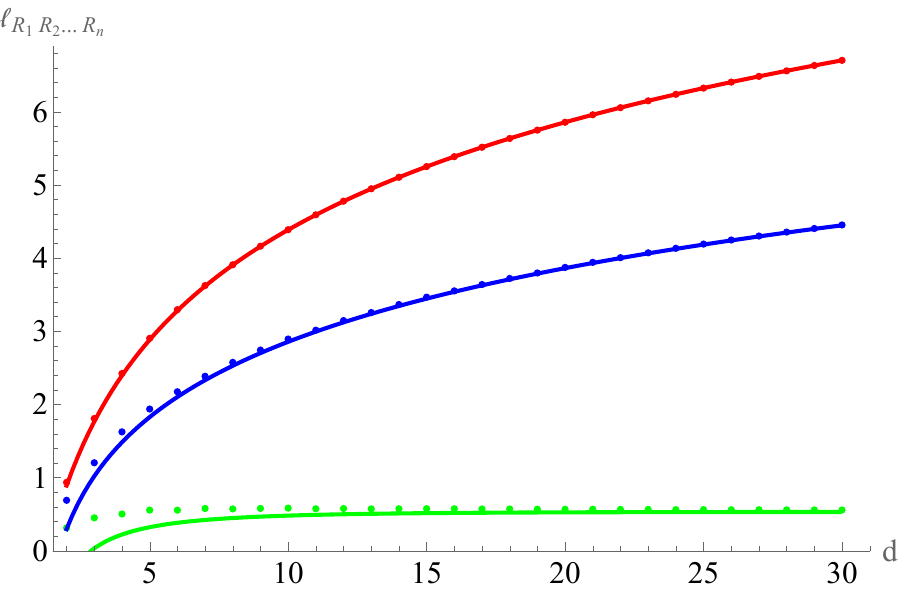}
	\end{adjustbox}
	\caption{$\ell_{A_1A_2\cdot\cdot\cdot A_n}$ for three-party (green), four-party (blue) and five-party (red) random states are plotted here with increasing dimension $d$. The dotted and the solid lines correspond to the numerical and the analytical results, respectively. }
	\label{randomplot}
\end{figure}

\textit{Random states.}---Here we focus on the analysis of multipartite entanglement for random pure states chosen from a Haar random distribution.
Utilizing the resolvent technique, the expression for reflected entropy between two subsystems $A_i$ and $A_j$ with dimensions $d_{A_i}$ and $d_{A_j}$ can be obtained as,
\begin{align}\label{refran}
	S_R(A_i:A_j)\approx-\sum_{r=1,2} p_r(q) \ln p_r(q)+p_2(q)\left(\ln d_{A_i}^2-\frac{d_{A_i}^2}{2 d_{A_j}^2}\right)
\end{align}
where $q=d_{A_i} d_{A_j}/d_{\overline{A_iA_j}}$ and $p_r(q)$ are the functions of $q$-Catalan number having different Hypergeometric behaviour depending on the signature of $q-1$  \cite{Akers:2021pvd}. Consequently, the analysis of L-entropy will exhibit distinct behaviours depending on the dimensions of the subsystems $A_i$, $A_j$ and $A_iA_j$. To investigate this property, we consider 3-party, 4-party, and 5-party systems. Note that, the expression for the reflected entropy also includes $\mathcal{O}(1/d_{A_i}^2)$, $\mathcal{O}(1/d_{A_j}^2)$ terms which we discard considering the dimensions of the subsystems to be large. Consequently, the evaluation of the L-entropy will be performed under the assumption that the dimensions of all the subsystems are all identical and large. Furthermore, any two subsystems yield identical bipartite entanglement, implying that the multipartite entanglement reduces to bipartite entanglement for any pair of subsystems. Leveraging this observation, we will focus on the computation of bipartite L-entropy for an arbitrary pair of subsystems. First, considering the tripartite random pure state where $q>1$, the bipartite L-entropy results as,
\begin{align}\label{Lenrandom3}
	\ell_{A_iA_j}=  \frac{1}{2} +\frac{2\log [d]-5}{2 d}+O(\frac{1}{d^2}),
\end{align}
in the large $d$ limit. Note that, the leading order results for the multipartite entanglement remain independent of $d$ and much smaller than their maximum value $\log[d]$, unlike the bipartite entanglement which shows maximality. Moving to the 4-party system, we find an interesting observation as $q=1$. The bipartite L-entropy can be obtained in the large $d$ limit as,
\begin{align}\label{random4p}
	\ell_{A_iA_j}=  2 x_0 \log[d]+y_0+O(\frac{1}{d^2})
\end{align}
where $x_0=p_1(1)\approx 0.720$, $y_0=-S_0(1)+p_2(1)/2\approx-0.453$ and $S_0(1)$ is the Shannon entropy $\sum_{r=1,2}-p_r(q) \ln p_r(q)$ for $q=1$. Interestingly, for $x_0<1$, the 4-party L-entropy in this phase scales as $\log[d]$, but it does not attain the maximum value $2\log[d]$. In the context of 5-party systems, we obtain $q<1$ where the $q$-Catalan number shows different behaviour compared to the 3-party and 4-party cases. Here, the bipartite L-entropy is expressed as,
\begin{align}\label{Lenrandom5}
	\ell_{A_iA_j}=2\log[d]-\frac{2\log (d^3/4)+1}{8 d}+O\left(\frac{1}{d^2}\right)
\end{align}  
up to the order of $\mathcal{O}(1/d^2)$. Interestingly, the multipartite L-entropy reaches the maximum value $2\log[d]$ unlike the 3-party and the 4-party cases. The bipartite L-entropy attains the same value as in \cref{Lenrandom5} for a maximally mixed state $\rho_{A_iA_j}=\mathcal{I}/{d}^2$. This value remains consistent across all bipartitions, corresponding to a particular class of states in quantum information theory known as 2-uniform states \cite{PhysRevA.90.022316}. In \cref{randomplot}, we plotted $\ell_{A_iA_j}$ both analytically utilizing \cref{Lenrandom3,random4p,Lenrandom5} and numerically using the expression in \cref{refran}. Interestingly, we observe that as the number of parties increases, the analytical and numerical results converge at progressively smaller values of $d$.

\begin{figure}[htp]
	\centering
	\subfigure[]{\includegraphics[width=.4\linewidth]{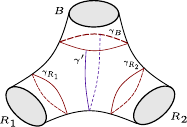}\label{mb_wh}}\quad
	\subfigure[]{\includegraphics[width=.5\linewidth]{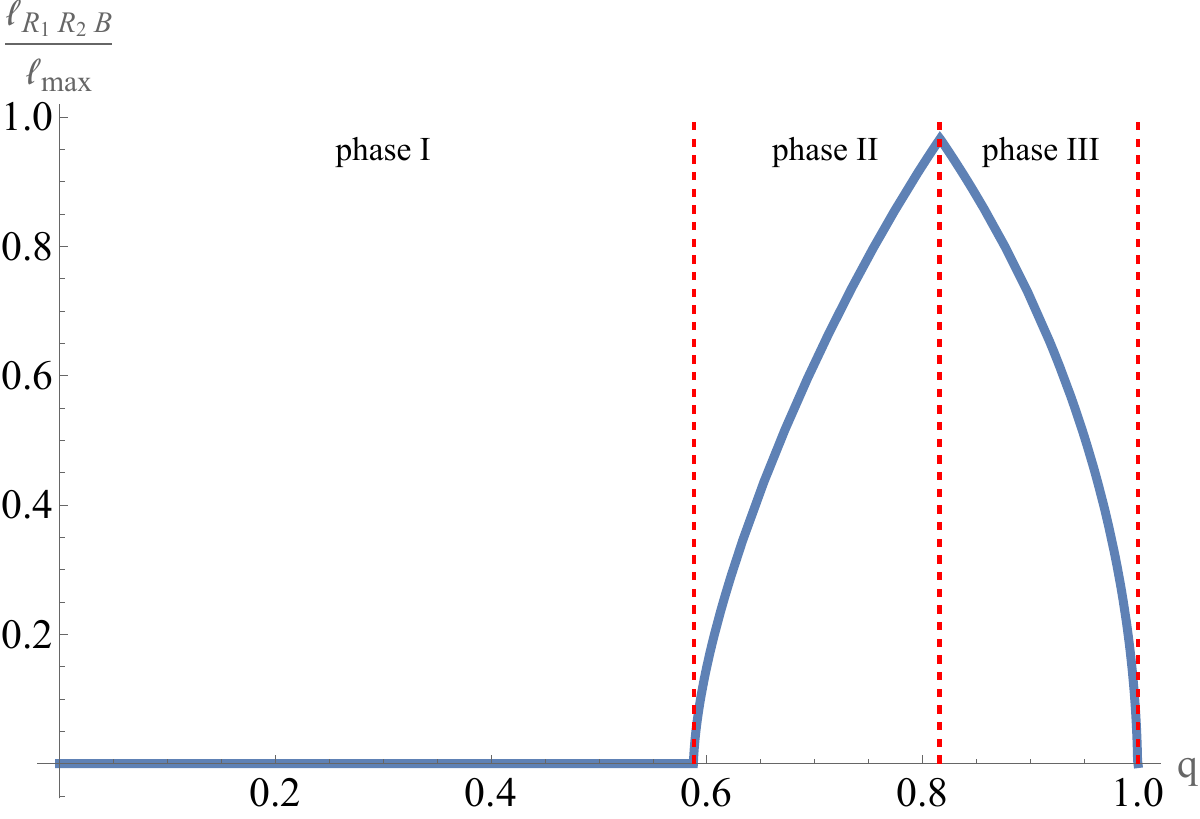}\label{l_mb_wh}}
\caption{(a) Three-boundary wormhole with black hole $B$ and radiation regions $R_1,R_2$. 
	(b) Page curve of tripartite L-entropy $\ell_{R_1R_2B}/\ell_{\max}$ versus 
	$q=\sqrt{2}\mathcal{A}_{R_1}/\mathcal{A}_{B_0}$. 
	Phases I--III correspond to 
	$\mathcal{A}_B>\mathcal{A}_{R_1}+\mathcal{A}_{R_2}$, 
	$\mathcal{A}_{R_1}<\mathcal{A}_B<\mathcal{A}_{R_1}+\mathcal{A}_{R_2}$, 
	and $\mathcal{A}_B<\mathcal{A}_{R_1}$, with $|R_1|=|R_2|$.}
\end{figure}

\textit{Holography.}---Now we delve into the computations of L-entropy to understand the characteristics of the multipartite entanglement in the context of $AdS/CFT$ correspondence. The L entropy can be computed in CFTs where it yields an universal information. See Supplementary Material for a detailed discussion \cite{suppl}. However, here we will focus on the geometric aspects of this quantity from a gravitational point of view. In $AdS/CFT$ correspondence, the entanglement entropy and the reflected entropy are given by the area of the bulk dual surfaces known as Ryu-Takayanagi (RT) surface \cite{Ryu:2006bv,Ryu:2006ef} and entanglement wedge cross section (EWCS) \cite{Takayanagi:2017knl}. Following these, the bulk dual of bipartite L-entropy for a pair of subsystems $A_i$ and $A_j$ at the boundary can be expressed as the difference of the area of a particular RT surfaces $\gamma_{A_i}$, $\gamma_{A_j}$ and the EWCS for the corresponding bipartition $\gamma^\prime_{A_iA_j}$ following \cref{bip_l_1}. 
\begin{align}
	\ell_{A_iA_j}=\frac{\text{Min}\left[\mathcal{A}(\gamma_{A_i}),\mathcal{A}(\gamma_{A_j})\right]-\mathcal{A}(\gamma^\prime_{A_iA_j})}{2G_N}.
\end{align}
We will utilize this construction for holographic bipartite L-entropy in the scenario of three-boundary wormhole following the model in \cite{Akers:2019nfi} where one of the boundaries can be thought of as a black hole $B$ and the other two as the radiation regions $R_1$ and $R_2$. In this specific model, all the boundaries together form a pure state which allows us to use tripartite L-entropy as a measure of genuine tripartite entanglement present in the system during the black hole evaporation. In Fig.~\ref{mb_wh}, $\gamma_{R_1},~\gamma_{R_2}$ and $\gamma_{B}$ are the HRT surfaces corresponding to the two radiation regions and the black hole respectively.  The entanglement wedge for the total radiation region ($R_1\cup R_2$) is the interior bulk region of the wormhole bounded by $\gamma_{R_1},~\gamma_{R_2}$ and $\gamma_{B}$. The plausible entanglement wedge cross sectional surfaces corresponding to this wedge geometry are $\gamma_{R_1},~\gamma_{R_2}$ and $\gamma^\prime$. 
Furthermore we invoke energy-entropy relation of the holography $S=2\pi\sqrt{cE/3}$ which relates the entanglement entropies of all the boundaries as $\mathcal{A}_{B_0}^2=\mathcal{A}_{R_1}^2+\mathcal{A}_{R_2}^2+\mathcal{A}_{B}^2$ with $\mathcal{A}_{B_0}$ being the initial horizon area of the black hole without any radiation. Consequently, we compute the bipartite L-entropy $\ell_{R_1R_2}$ and the same for other bipartitions considering separate entanglement wedge and EWCS. Combining all these bipartite L-entropy, we analyze the evolution of the tripartite L-entropy for the whole system in \cref{l_mb_wh} with increasing size of radiation region where $R_1$ and $R_2$ are considered to be of the same size for simplicity. Interestingly, we observe vanishing tripartite L-entropy and thus the absence of genuine tripartite entanglement (phase I) until the Page time when the whole radiation region $R_1 \cup R_2$ gets access to the island and $\mathcal{A}_{B}=\mathcal{A}_{R_1}+\mathcal{A}_{R_2}$ \cite{Akers:2019nfi}. Subsequently, the tripartite L-entropy shows a significant increase (phase II) until a maximum value $\ell_{R_1R_2B}=\ell_{max}$ when $\mathcal{S}_{R_1}=\mathcal{S}_{R_2}=\mathcal{S}_{B}$. Here, the tripartite L-entropy becomes exactly equal to the entanglement entropy of either the black hole or the radiation subsystems, which perfectly aligns with the expected bound. However, after obtaining this maximum value, the genuine multipartite entanglement decreases (phase III) with further evaporation of the black hole. Note that, in this region, the black hole becomes smaller than the total radiation region $\mathcal{A}_{B}<\mathcal{A}_{R_1}+\mathcal{A}_{R_2}$. Finally the tripartite L-entropy becomes zero when the black hole evaporates completely which leaves the total system as a bipartite pure state $R_1\cup R_2$.

\textit{GME at finite temperature.}---To extend our multipartite entanglement measure to finite temperatures, we employ thermal pure quantum (TPQ) states~\cite{Sugiura:2013pla}, which reproduce thermal expectation values while remaining pure. The canonical TPQ state is defined as
\begin{align}
	|\Psi(\beta)\rangle \equiv e^{-{\beta\over 2}H}|\psi\rangle ,
\end{align}
where $H$ is the system Hamiltonian and $|\psi\rangle$ is a random state, yielding
\begin{align}
	\frac{\overline{\langle \Psi_\beta | \cO |\Psi_\beta \rangle}}
	{\overline{\langle \Psi_\beta | \Psi_\beta \rangle}}
	= \frac{1}{Z(\beta)} \tr(\cO e^{-\beta H}) .
\end{align}
We generalize this construction to multipartite systems by defining the multipartite thermal pure quantum (MTPQ) state
\begin{align}
	|\Psi_\alpha\rangle \equiv \bigotimes_{i=1}^n e^{-{\alpha\over 2}H^{(k)}} |\psi\rangle ,
\end{align}
which we interpret as a thermal 2-uniform state, characterized by factorized two-party reduced density matrices
\begin{align}
	\rho_{A_{j_1}A_{j_k}} = \rho_{\text{\tiny th}}(\beta_{j_1}) \otimes \rho_{\text{\tiny th}}(\beta_{j_k}) .
\end{align}
The corresponding bipartite L-entropy satisfies
\begin{align}\label{Ltherm2uni}
	\ell_{A_iA_j} = 2\min\!\big(S_{\text{\tiny th}}(\beta_i),S_{\text{\tiny th}}(\beta_j)\big) .
\end{align}

\begin{figure}[!t]
	\centering
	\begin{adjustbox}{center}
		\includegraphics[width=.8\linewidth]{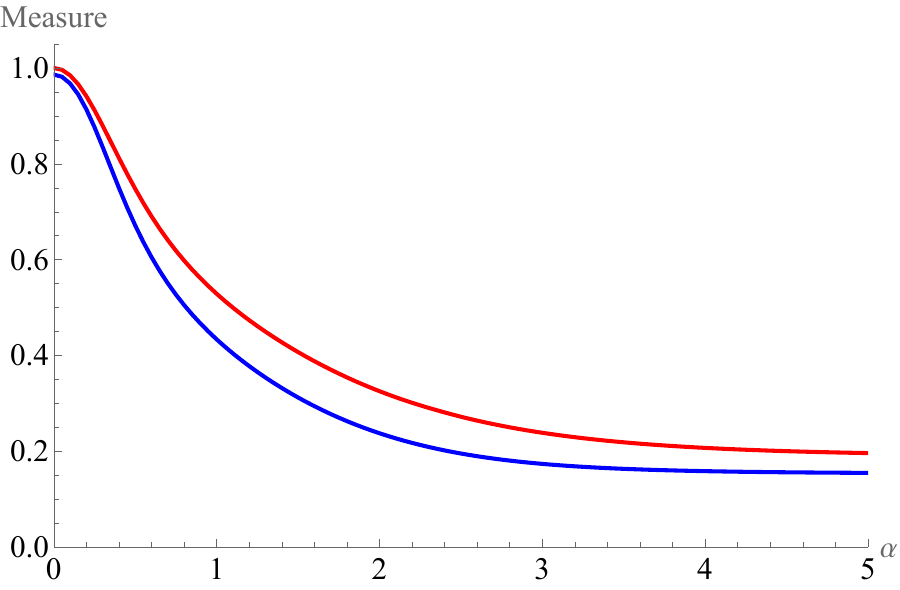}
	\end{adjustbox}
	\caption{$\ell_{A_iA_j}$ (blue) and its value for thermal 2-uniform state   given by \cref{Ltherm2uni} (red) are plotted with increasing $\alpha$ for multi-copy SYK model each involving 5-parties.}
	\label{finitetemp}
\end{figure}

\textit{MTPQ in SYK model.}---For the multi-copy SYK model, numerical results shown in Fig.~\ref{finitetemp} demonstrate that the L-entropy for five parties closely follows the thermal 2-uniform prediction.

\textit{Conclusion.}---In this letter, we introduced a new measure to characterize genuine multipartite entanglement, applicable to $n \leq 5$ pure states in finite-dimensional quantum systems, quantum field theories, and holography. We demonstrated that our measure satisfies all fundamental properties required by quantum information theory. Notably, it attains its maximal value for GHZ states in three-party systems and for $k \geq 2$ uniform states in four- and five-party systems.We applied our measure to random multipartite states and found that, to leading order in the Hilbert space dimension of individual subsystems, five-party random states exhibit maximal entanglement, whereas three- and four-party states do not. Additionally, we computed our measure in a holographic CFT dual to a three-boundary wormhole, uncovering an analogue of the Page curve for multipartite entanglement. Extending our analysis, we proposed a multipartite generalization of the thermal pure quantum (TPQ) state and evaluated our measure in such a state within a multi-copy SYK model. This revealed a thermal two-uniform behavior, extending the concept of 2-uniform states to finite temperatures.

\textit{Acknowledgement.}---The authors would like to thank Chris Akers, Eunok Bae, Minjin Choi, Girish Kulkarni, Hyukjoon Kwon, Takato Mori, and Kornikar Sen for their valuable discussions and remarks. The research work of JKB is supported by the Brain Pool program funded by the Ministry of Science and ICT through the National Research Foundation of Korea (RS-2024-00445164). V.M. and J.Y. was supported by the National Research Foundation of Korea (NRF) grant funded by the Korean government (MSIT) (RS-2022-NR069038) and by the Brain Pool program funded by the Ministry of Science and ICT through the National Research Foundation of Korea (RS-2023-00261799). 

\bibliography{GME_prl_rev_nofoottex}

@article{newlentropy,
		title={Probing the Hierarchy of Genuine Multipartite Entanglement with Generalized Latent Entropy}, 
		author={Byoungjoon Ahn and Jaydeep Kumar Basak and Keun-Young Kim and Gwon Bin Koo and Vinay Malvimat and Junggi Yoon},
		year={2025},
		eprint={2510.19922},
		archivePrefix={arXiv},
		primaryClass={hep-th},
		url={https://arxiv.org/abs/2510.19922}, 
	}

@misc{suppl,
	note = {See the Supplemental Material for further details}}

@article{PhysRevLett.113.110501,
  title = {Strong Monogamy Conjecture for Multiqubit Entanglement: The Four-Qubit Case},
  author = {Regula, Bartosz and Di Martino, Sara and Lee, Soojoon and Adesso, Gerardo},
  journal = {Phys. Rev. Lett.},
  volume = {113},
  issue = {11},
  pages = {110501},
  numpages = {6},
  year = {2014},
  month = {Sep},
  publisher = {American Physical Society},
  doi = {10.1103/PhysRevLett.113.110501},
  url = {https://link.aps.org/doi/10.1103/PhysRevLett.113.110501}
}

@article{PhysRevA.80.032330,
  title = {Multipartite bound entanglement and multisetting Bell inequalities},
  author = {Chi, Dong Pyo and Jeong, Kabgyun and Kim, Taewan and Lee, Kyungjin and Lee, Soojoon},
  journal = {Phys. Rev. A},
  volume = {80},
  issue = {3},
  pages = {032330},
  numpages = {4},
  year = {2009},
  month = {Sep},
  publisher = {American Physical Society},
  doi = {10.1103/PhysRevA.80.032330},
  url = {https://link.aps.org/doi/10.1103/PhysRevA.80.032330}
}

@article{PhysRevA.90.022316,
  title = {Genuinely multipartite entangled states and orthogonal arrays},
  author = {Goyeneche, Dardo and \ifmmode \dot{Z}\else \.{Z}\fi{}yczkowski, Karol},
  journal = {Phys. Rev. A},
  volume = {90},
  issue = {2},
  pages = {022316},
  numpages = {18},
  year = {2014},
  month = {Aug},
  publisher = {American Physical Society},
  doi = {10.1103/PhysRevA.90.022316},
  url = {https://link.aps.org/doi/10.1103/PhysRevA.90.022316}
}

@article{Li:2021hhj,
    author = "Li, Yinfei and Shang, Jiangwei",
    title = "{Geometric mean of bipartite concurrences as a genuine multipartite entanglement measure}",
    eprint = "2112.10509",
    archivePrefix = "arXiv",
    primaryClass = "quant-ph",
    doi = "10.1103/PhysRevResearch.4.023059",
    journal = "Phys. Rev. Res.",
    volume = "4",
    number = "2",
    pages = "023059",
    year = "2022"
}

@article{Gadde:2024jfi,
    author = "Gadde, Abhijit and Jain, Shraiyance and Kulkarni, Harshal",
    title = "{Multi-partite entanglement monotones}",
    eprint = "2406.17447",
    archivePrefix = "arXiv",
    primaryClass = "quant-ph",
    reportNumber = "TIFR/TH/24-10",
    month = "6",
    year = "2024"
}

@article{Sugiura:2013pla,
	archiveprefix = {arXiv},
	author = {Sugiura, Sho and Shimizu, Akira},
	doi = {10.1103/PhysRevLett.111.010401},
	eprint = {1302.3138},
	journal = {Phys. Rev. Lett.},
	number = {1},
	pages = {010401},
	primaryclass = {cond-mat.stat-mech},
	title = {{Canonical Thermal Pure Quantum State}},
	volume = {111},
	year = {2013}}

@article{Ma:2023ecg,
	archiveprefix = {arXiv},
	author = {Ma, Mengru and Li, Yinfei and Shang, Jiangwei},
	eprint = {2309.09459},
	month = {9},
	primaryclass = {quant-ph},
	title = {{Multipartite entanglement measures: a review}},
	year = {2023}}

@article{PhysRevLett.127.040403,
	author = {Xie, Songbo and Eberly, Joseph H.},
	doi = {10.1103/PhysRevLett.127.040403},
	issue = {4},
	journal = {Phys. Rev. Lett.},
	month = {Jul},
	numpages = {6},
	pages = {040403},
	publisher = {American Physical Society},
	title = {Triangle Measure of Tripartite Entanglement},
	url = {https://link.aps.org/doi/10.1103/PhysRevLett.127.040403},
	volume = {127},
	year = {2021}}

@article{Akers:2019gcv,
    author = "Akers, Chris and Rath, Pratik",
    title = "{Entanglement Wedge Cross Sections Require Tripartite Entanglement}",
    eprint = "1911.07852",
    archivePrefix = "arXiv",
    primaryClass = "hep-th",
    doi = "10.1007/JHEP04(2020)208",
    journal = "JHEP",
    volume = "04",
    pages = "208",
    year = "2020"
}

@article{Gadde:2022cqi,
    author = "Gadde, Abhijit and Krishna, Vineeth and Sharma, Trakshu",
    title = "{New multipartite entanglement measure and its holographic dual}",
    eprint = "2206.09723",
    archivePrefix = "arXiv",
    primaryClass = "hep-th",
    reportNumber = "TIFR/TH/22-34",
    doi = "10.1103/PhysRevD.106.126001",
    journal = "Phys. Rev. D",
    volume = "106",
    number = "12",
    pages = "126001",
    year = "2022"
}

@article{Cleve:1999qg,
    author = "Cleve, Richard and Gottesman, Daniel and Lo, Hoi-Kwong",
    title = "{How to share a quantum secret}",
    eprint = "quant-ph/9901025",
    archivePrefix = "arXiv",
    reportNumber = "LA-UR-98-5842, LAUR98-5842",
    doi = "10.1103/PhysRevLett.83.648",
    journal = "Phys. Rev. Lett.",
    volume = "83",
    pages = "648--651",
    year = "1999"
}

@article{Shi:2025gyu,
    author = "Shi, Hai-Long and Gan, Li and Zhang, Kun and Wang, Xiao-Hui and Yang, Wen-Li",
    title = "{Quantum Charging Advantage from Multipartite Entanglement}",
    eprint = "2503.02667",
    archivePrefix = "arXiv",
    primaryClass = "quant-ph",
    month = "3",
    year = "2025"
}

@article{briegel2009measurement,
  title={Measurement-based quantum computation},
  author={Briegel, Hans J and Browne, David E and D{\"u}r, Wolfgang and Raussendorf, Robert and Van den Nest, Maarten},
  journal={Nature Physics},
  volume={5},
  number={1},
  pages={19--26},
  year={2009},
  publisher={Nature Publishing Group UK London}
}

@article{Wong:2022mnv,
    author = "Wong, Gabriel and Raussendorf, Robert and Czech, Bartlomiej",
    title = "{The Gauge Theory of Measurement-Based Quantum Computation}",
    eprint = "2207.10098",
    archivePrefix = "arXiv",
    primaryClass = "hep-th",
    doi = "10.22331/q-2024-07-04-1397",
    journal = "Quantum",
    volume = "8",
    pages = "1397",
    year = "2024"
}

@article{Iizuka:2024pzm,
    author = "Iizuka, Norihiro and Lin, Simon and Nishida, Mitsuhiro",
    title = "{Black hole multi-entropy curves \textemdash{} secret entanglement between Hawking particles}",
    eprint = "2412.07549",
    archivePrefix = "arXiv",
    primaryClass = "hep-th",
    doi = "10.1007/JHEP03(2025)037",
    journal = "JHEP",
    volume = "03",
    pages = "037",
    year = "2025"
}

@article{Hillery:1998yq,
    author = "Hillery, Mark and Buzek, Vladimir and Berthiaume, Andre",
    title = "{Quantum secret sharing}",
    eprint = "quant-ph/9806063",
    archivePrefix = "arXiv",
    doi = "10.1103/PhysRevA.59.1829",
    journal = "Phys. Rev. A",
    volume = "59",
    pages = "1829",
    year = "1999"
}

@article{Choi:2022lge,
    author = "Choi, Minjin and Bae, Eunok and Lee, Soojoon",
    title = "{Genuine multipartite entanglement measures based on multi-party teleportation capability}",
    eprint = "2211.15986",
    archivePrefix = "arXiv",
    primaryClass = "quant-ph",
    doi = "10.1038/s41598-023-42052-x",
    journal = "Sci. Rep.",
    volume = "13",
    number = "1",
    pages = "15013",
    year = "2023"
}

@article{Ge:2022sqp,
    author = "Ge, Xiaozhen and Liu, Lijun and Cheng, Shuming",
    title = "{Tripartite entanglement measure under local operations and classical communication}",
    eprint = "2210.06700",
    archivePrefix = "arXiv",
    primaryClass = "quant-ph",
    doi = "10.1103/PhysRevA.107.032405",
    journal = "Phys. Rev. A",
    volume = "107",
    number = "3",
    pages = "032405",
    year = "2023"
}

@article{Horodecki:2024bgc,
    author = "Horodecki, Pawel and Rudnicki, \L{}ukasz and \.Zyczkowski, Karol",
    title = "{Multipartite entanglement}",
    eprint = "2409.04566",
    archivePrefix = "arXiv",
    primaryClass = "quant-ph",
    month = "9",
    year = "2024"
}

@article{Dutta:2019gen,
	archiveprefix = {arXiv},
	author = {Dutta, Souvik and Faulkner, Thomas},
	doi = {10.1007/JHEP03(2021)178},
	eprint = {1905.00577},
	journal = {JHEP},
	pages = {178},
	primaryclass = {hep-th},
	title = {{A canonical purification for the entanglement wedge cross-section}},
	volume = {03},
	year = {2021}}

@article{Akers:2019nfi,
	archiveprefix = {arXiv},
	author = {Akers, Chris and Engelhardt, Netta and Harlow, Daniel},
	doi = {10.1007/JHEP08(2020)032},
	eprint = {1910.00972},
	journal = {JHEP},
	pages = {032},
	primaryclass = {hep-th},
	title = {{Simple holographic models of black hole evaporation}},
	volume = {08},
	year = {2020}}

@article{Takayanagi:2017knl,
	archiveprefix = {arXiv},
	author = {Takayanagi, Tadashi and Umemoto, Koji},
	doi = {10.1038/s41567-018-0075-2},
	eprint = {1708.09393},
	journal = {Nature Phys.},
	number = {6},
	pages = {573--577},
	primaryclass = {hep-th},
	reportnumber = {YITP-17-89, IPMU17-0115},
	title = {{Entanglement of purification through holographic duality}},
	volume = {14},
	year = {2018}}

@article{Hayden:2023yij,
	archiveprefix = {arXiv},
	author = {Hayden, Patrick and Lemm, Marius and Sorce, Jonathan},
	eprint = {2302.10208},
	month = {2},
	primaryclass = {hep-th},
	reportnumber = {MIT-CTP/5532},
	title = {{Reflected entropy is not a correlation measure}},
	year = {2023}}

@article{Zou:2020bly,
	archiveprefix = {arXiv},
	author = {Zou, Yijian and Siva, Karthik and Soejima, Tomohiro and Mong, Roger S. K. and Zaletel, Michael P.},
	doi = {10.1103/PhysRevLett.126.120501},
	eprint = {2011.11864},
	journal = {Phys. Rev. Lett.},
	number = {12},
	pages = {120501},
	primaryclass = {quant-ph},
	title = {{Universal tripartite entanglement in one-dimensional many-body systems}},
	volume = {126},
	year = {2021}}

@article{Akers:2021pvd,
	archiveprefix = {arXiv},
	author = {Akers, Chris and Faulkner, Thomas and Lin, Simon and Rath, Pratik},
	doi = {10.1007/JHEP05(2022)162},
	eprint = {2112.09122},
	journal = {JHEP},
	pages = {162},
	primaryclass = {hep-th},
	title = {{Reflected entropy in random tensor networks}},
	volume = {05},
	year = {2022}}

@article{Plenio:2007zz,
	archiveprefix = {arXiv},
	author = {Plenio, Martin B. and Virmani, Shashank},
	eprint = {quant-ph/0504163},
	journal = {Quant. Inf. Comput.},
	pages = {1--51},
	title = {{An Introduction to entanglement measures}},
	volume = {7},
	year = {2007}}

@article{Hayden:2021gno,
	archiveprefix = {arXiv},
	author = {Hayden, Patrick and Parrikar, Onkar and Sorce, Jonathan},
	doi = {10.1007/JHEP10(2021)047},
	eprint = {2107.00009},
	journal = {JHEP},
	pages = {047},
	primaryclass = {hep-th},
	title = {{The Markov gap for geometric reflected entropy}},
	volume = {10},
	year = {2021}}

@article{Ryu:2006ef,
	archiveprefix = {arXiv},
	author = {Ryu, Shinsei and Takayanagi, Tadashi},
	doi = {10.1088/1126-6708/2006/08/045},
	eprint = {hep-th/0605073},
	journal = {JHEP},
	pages = {045},
	reportnumber = {NSF-KITP-06-31, KUNS-2021},
	title = {{Aspects of Holographic Entanglement Entropy}},
	volume = {08},
	year = {2006}}

@article{Ryu:2006bv,
	archiveprefix = {arXiv},
	author = {Ryu, Shinsei and Takayanagi, Tadashi},
	doi = {10.1103/PhysRevLett.96.181602},
	eprint = {hep-th/0603001},
	journal = {Phys. Rev. Lett.},
	pages = {181602},
	reportnumber = {NSF-KITP-06-11},
	title = {{Holographic derivation of entanglement entropy from AdS/CFT}},
	volume = {96},
	year = {2006}}

@article{Dur:2000zz,
    author = "Dur, W. and Vidal, G. and Cirac, J. I.",
    title = "{Three qubits can be entangled in two inequivalent ways}",
    eprint = "quant-ph/0005115",
    archivePrefix = "arXiv",
    doi = "10.1103/PhysRevA.62.062314",
    journal = "Phys. Rev. A",
    volume = "62",
    pages = "062314",
    year = "2000"
}

@article{Coffman:1999jd,
	archiveprefix = {arXiv},
	author = {Coffman, Valerie and Kundu, Joydip and Wootters, William K.},
	doi = {10.1103/PhysRevA.61.052306},
	eprint = {quant-ph/9907047},
	journal = {Phys. Rev. A},
	pages = {052306},
	title = {{Distributed entanglement}},
	volume = {61},
	year = {2000}}

@article{Sabin:2008naa,
		author = "Sab{\'\i}n, Carlos and Garci{\-}a-Alcaine, Guillermo",
		title = "{A classification of entanglement in three-qubit systems}",
		eprint = "0707.1780",
		archivePrefix = "arXiv",
		primaryClass = "quant-ph",
		doi = "10.1140/epjd/e2008-00112-5",
		journal = "Eur. Phys. J. D",
		volume = "48",
		number = "3",
		pages = "435--442",
		year = "2008"
	}

@article{Scott:2004vaq,
		author = "Scott, A. J.",
		title = "{Multipartite entanglement, quantum-error-correcting codes, and entangling power of quantum evolutions}",
		eprint = "quant-ph/0310137",
		archivePrefix = "arXiv",
		doi = "10.1103/PhysRevA.69.052330",
		journal = "Phys. Rev. A",
		volume = "69",
		number = "5",
		pages = "052330",
		year = "2004"
	}

@article{PhysRevA.86.052335,
		title = {Absolute maximal entanglement and quantum secret sharing},
		author = {Helwig, Wolfram and Cui, Wei and Latorre, Jos\'e Ignacio and Riera, Arnau and Lo, Hoi-Kwong},
		journal = {Phys. Rev. A},
		volume = {86},
		issue = {5},
		pages = {052335},
		numpages = {5},
		year = {2012},
		month = {Nov},
		publisher = {American Physical Society},
		doi = {10.1103/PhysRevA.86.052335},
		url = {https://link.aps.org/doi/10.1103/PhysRevA.86.052335}
	}

@article{PhysRevA.81.012308,
		title = {Channel capacities versus entanglement measures in multiparty quantum states},
		author = {Sen(De), Aditi and Sen, Ujjwal},
		journal = {Phys. Rev. A},
		volume = {81},
		issue = {1},
		pages = {012308},
		numpages = {6},
		year = {2010},
		month = {Jan},
		publisher = {American Physical Society},
		doi = {10.1103/PhysRevA.81.012308},
		url = {https://link.aps.org/doi/10.1103/PhysRevA.81.012308}
	}

@article{Xie:2023dzw,
		author = "Xie, Songbo and Younis, Daniel and Mei, Yuhan and Eberly, Joseph H.",
		title = "{Multipartite Entanglement: A Journey through Geometry}",
		eprint = "2304.03281",
		archivePrefix = "arXiv",
		primaryClass = "quant-ph",
		doi = "10.3390/e26030217",
		journal = "Entropy",
		volume = "26",
		number = "3",
		pages = "217",
		year = "2024"
	}

@article{Ma:2011yon,
		author = "Ma, Zhi-Hao and Chen, Zhi-Hua and Chen, Jing-Ling and Spengler, Christoph and Gabriel, Andreas and Huber, Marcus",
		title = "{Measure of genuine multipartite entanglement with computable lower bounds}",
		eprint = "1101.2001",
		archivePrefix = "arXiv",
		primaryClass = "quant-ph",
		doi = "10.1103/PhysRevA.83.062325",
		journal = "Phys. Rev. A",
		volume = "83",
		number = "6",
		pages = "062325",
		year = "2011"
	}

\onecolumngrid
\newpage
\begin{center}
  \textbf{\large Supplementary Material}\\[.2cm]
\end{center}
\setcounter{equation}{0}
\setcounter{figure}{0}
\setcounter{table}{0}
\setcounter{page}{1}
\setcounter{section}{0}
\renewcommand{\theequation}{S\arabic{equation}}
\renewcommand{\thefigure}{S\arabic{figure}}
\renewcommand{\bibnumfmt}[1]{[S#1]}
\renewcommand{\thesection}{Supp.\arabic{section}}

\section{Monotonicity}\label{sup_mono}

A pure state entanglement monotone is defined as a local unitary invariant function which is concave under any local operation \cite{Gadde:2024jfi}

\begin{align}\label{mono1}
	f(|\psi\rangle_{\textrm{\tiny{ABC}}}) \geq \sum_i p_i f\left(\left|\psi_i\right\rangle_{\textrm{\tiny{ABC}}}\right)
\end{align}
 $|\psi_i\rangle_{\textrm{\tiny{ABC}}}$ in the above equation denotes the states obtained  post a one party local operation such that the state $\rho=\ket{\psi}\bra{\psi}$ is converted into $\Lambda(\rho)$ as follows
\begin{align}\label{Lam}
	\Lambda(\rho)=\sum_i p_i|\psi_i \rangle\langle\psi_i|, \quad p_i:=|| E_i^{(A) }\ket{\psi}||^ 2, \quad \ket{\psi_i}:=E_i^{(A)}|\psi\rangle / \sqrt{p_i}
\end{align}
 $E_i^{A}$ correspond to a trace preserving operation on subsystem $A$
\begin{align}
 \sum_i E_i^{\dagger (A)} E_i^{(A)}=\mathbb{I}
\end{align}

Consider the state \(\ket{\phi}\), which serves as a purification of the reduced density matrix \(\rho_{\textrm{\tiny{AB}}}\). This reduced density matrix is obtained by tracing out subsystem \(C\) from the original state \(\ket{\psi}\). Likewise, we define \(\ket{\phi_i}\) as the purification of the reduced density matrix \(\rho_{i,\textrm{\tiny{AB}}} = \mathrm{Tr}_C(\ket{\psi_i})\). The state \(\ket{\phi}\) resides in the Hilbert space \(\mathcal{H}_{AB} \otimes \mathcal{H}_{\tilde{A}}\), where the auxiliary space \(\mathcal{H}_{\tilde{A}}\) must be at least as large in dimension as the rank of \(\rho_{\textrm{\tiny{AB}}}\).  Since all possible purifications of the bipartite system—including the original state \(\ket{\psi}\)—are connected via such transformations and the function \(f\) is a local unitary invariant,  it follows that \(f\) takes the same value for any of these purifications.

\begin{align}
    f(\ket{\phi})=f(\ket{\psi}),\quad  f(\ket{\phi_i})=f(\ket{\psi_i})
\end{align}
Therefore,  \cref{mono1} may equally well be expressed through the canonically purifications as follows
\begin{align}
	f(\ket{\phi}) \geq \sum_i p_i f(\ket{\phi_i})
\end{align}
We can now utilize the above result in the context of our measure which  is obtained through the canonical purification, where \(\ket{\phi}\) resides in the Hilbert space \(\mathcal{H}_{AB} \otimes \mathcal{H}_{A^*B^*}\). Here, the auxiliary Hilbert space \(\mathcal{H}_{A^*B^*}\) has a dimension exactly equal to the rank of the reduced density matrix \(\rho_{\textrm{\tiny{AB}}}\).  
In the context of canonical purification, our next step is to demonstrate that

\begin{align}
	f(\ket{\sqrt{\rho_{\textrm{\tiny{AB}}}}}) \geq \sum_i p_i  f(\ket{\sqrt{\rho_{\textrm{i,\tiny{AB}}}}})
\end{align}
In the above equation \(\ket{\sqrt{\rho_{\textrm{\tiny{AB}}}}}\) and \(\ket{\sqrt{\rho_{\textrm{i,\tiny{AB}}}}}\) correspond to the canonical purifications of the reduced density matrices \(\rho_{\textrm{\tiny{AB}}}\) and \(\rho_{i,\textrm{\tiny{AB}}}\), respectively.  
Observe that the action of the map \(\Lambda\) on \(\rho\) in \cref{Lam} can equivalently be interpreted as the effect of the map \(\tilde{\Lambda}\) on the canonically purified state \(\ket{\sqrt{\rho_{\textrm{\tiny{AB}}}}}\). This transformation results in the states \(\ket{\sqrt{\rho_{\textrm{i,\tiny{AB}}}}}\) with probabilities

\begin{align}
	\tilde{\Lambda}( \ket{\sqrt{\rho_{\textrm{\tiny{AB}}}}})=\sum_i p_i \ket{\sqrt{\rho_{\textrm{i,\tiny{AB}}}}}\bra{\sqrt{\rho_{\textrm{i,\tiny{AB}}}}}.
\end{align}
We now aim to show that \(\tilde{\Lambda}\) corresponds simply to a local operation \(\Lambda\) acting on \(A\) and its reflected counterpart \(A^*\). To illustrate this, let us examine the reduced density matrix \(\rho_{i,\textrm{\tiny{AB}}}\) that results from applying a local operation \(E_i\) on \(A\) alone.
\begin{align}\label{rhoi}
	\rho_{i,AB}&=Tr_{C}(\rho_{\psi_i})\notag\\
	&=\frac{E_i^{(A)}Tr_C(\rho_{\psi})E_i^{\dagger(A)}}{p_i}\\
	&=\frac{E_i^{(A)}\rho_{AB}E_i^{\dagger(A)}}{p_i}
\end{align}
Since \(E_i^{(A)}\) acts only on subsystem \(A\), we can interchange the order of the operation and the trace over \(C\) in the second line.  Next, we express the reduced density matrix \(\rho_{\textrm{\tiny{AB}}}\) as a sum of pure states:
\begin{align}\label{redexp}
	\rho_{AB}=\sum_i q_i \ket{\lambda_i}\bra{\lambda_i}
\end{align}
The corresponding canonical purification is given by
\begin{align}
	\ket{\sqrt{\rho_{\textrm{\tiny{AB}}}}}=\sum_i \sqrt{q_i} \ket{\lambda_i}\ket{\lambda_i}
\end{align}
We may now utilize \cref{rhoi} and \cref{redexp} to obtain
\begin{align}
	\rho_{i,AB}=\sum_j q_j \frac{E_i^{(A)}}{\sqrt{p_i}}\ket{\lambda_j}\bra{\lambda_j}\frac{E_i^{\dagger(A)}}{\sqrt{p_i}}
\end{align}   
The corresponding canonical purification therefore is given by
\begin{align}
	\ket{\sqrt{\rho_{i,\textrm{\tiny{AB}}}}}&=\sum_j \sqrt{q_j} \frac{E^{(A)}_i\otimes E^{(A^*)}_i}{p_i}\ket{\lambda_j}\ket{\lambda_j}\notag\\
	\ket{\sqrt{\rho_{i,\textrm{\tiny{AB}}}}}& =\frac{E^{(A)}_i}{\sqrt{p_i}}\otimes \frac{E^{(A^*)}_i}{\sqrt{p_i}}\ket{\sqrt{\rho_{\textrm{\tiny{AB}}}}}&
\end{align}  
This conclusively shows that a local operation on \(A\) in the original state induces a corresponding local operation on both \(A\) and its reflected counterpart \(A^*\) in the canonically purified states.  Having established this, our next step is to demonstrate that the \(L\)-entropy satisfies the monotonicity property under local operations on \(A\) and \(A^*\). To do so, we invoke a well-known result from quantum information theory: the mutual information is non-increasing under local quantum operations, a property commonly referred to as the \textit{data processing inequality}.

\begin{align}
	I(A:A^*)_{\ket{\sqrt{\rho_{\textrm{i,\tiny{AB}}}}}} \leq I(A:A^*)_{\sqrt{\rho_{AB}} }\nonumber\\
	I(B:B^*)_{\ket{\sqrt{\rho_{\textrm{i,\tiny{AB}}}}}} \leq I(B:B^*)_{\sqrt{\rho_{AB}} }  
\end{align}
This further implies that
\begin{align}
	Min[ I(A:A^*),I(B:B^*)]_{\ket{\sqrt{\rho_{\textrm{i,\tiny{AB}}}}}}\leq Min[I(A:A^*),I(B:B^*)]_{\sqrt{\rho_{AB}}}\nonumber\\
\end{align}
Therefore, we obtain the following result
\begin{align}
	\ell_{AB}\geq \sum_i p_i \ell_{i,AB}\nonumber\\
\end{align}
where \(\ell_{\textrm{\tiny{AB}}}\) and \(\ell_{\textrm{i,\tiny{AB}}}\) denote the \(L\)-entropies associated with the canonically purified states \(\ket{\sqrt{\rho_{\textrm{\tiny{AB}}}}}\) and \(\ket{\sqrt{\rho_{\textrm{i,\tiny{AB}}}}}\), respectively.  Notice that in the proof above, we analyzed monotonicity with respect to a local operation on \(A\). The same reasoning applies to local operations on \(B\), while a local operation on \(C\) leaves \(\ell_{\textrm{\tiny{AB}}}\) unchanged.  
Therefore, we have established the proof that \(\ell_{\textrm{\tiny{AB}}}\) is a pure-state entanglement monotone.

The same reasoning applies to the purifications of \(\rho_{\textrm{\tiny{BC}}}\) and \(\rho_{\textrm{\tiny{AC}}}\), which confirms that \(\ell_{\textrm{\tiny{BC}}}\) and \(\ell_{\textrm{\tiny{AC}}}\) also qualify as entanglement monotones.  
Furthermore, as established in \cite{Ge:2022sqp}, the product of entanglement monotones, when appropriately exponentiated, retains its monotonicity. This ensures that \(\ell_{\textrm{\tiny{ABC}}}\) is likewise an entanglement monotone.  Additionally, the definition of the multipartite \(L\)-entropy satisfies condition (3) since each \(\ell_{A_i A_j}\) is an entanglement monotone, and their geometric mean preserves this property, as shown in \cite{Li:2021hhj}.  Notably, in our previous proof, we considered the subsystem \(\overline{A_i A_j}\) as \(C\). However, even when \(\overline{A_i A_j}\)  involves multiple parties the proof is considerably  straightforward as we will now demonstrate,
\begin{align}
\ell_{A_1A_2\cdot\cdot\cdot A_n}(\ket{\psi}_{A_1A_2\cdot\cdot\cdot A_n})&=\left(\prod_{i=1}^{n}\prod_{j=i+1}^{n}\ell_{A_iA_j}(\ket{\psi}_{A_iA_j\overline{A_iA_j}})\right)^{\frac{2}{n(n-1)}}\nonumber\\&\geq \left(\prod_{i=1}^{n}\prod_{j=i+1}^{n}\sum_k p_k \ell_{A_{i} A_{j}}\left(\ket{\psi_k}_{A_iA_j\overline{A_iA_j}}\right)\right)^{\frac{2}{n(n-1)}}. \nonumber\\
& \geq \sum_k p_k\left(\prod_{i=1}^{n}\prod_{j=i+1}^{n} \ell_{A_{i} A_{j}}\left(\ket{\psi_k}_{A_iA_j\overline{A_iA_j}}\right)\right)^{\frac{2}{n(n-1)}}.
\end{align}
An important point to highlight is that \cite{Hayden:2023yij} demonstrated that for a generic state \(\rho_{\textrm{\tiny{ABC}}}\), reflected entropy may fail to satisfy monotonicity under a partial trace. This raises a natural concern about whether a measure derived from reflected entropy can truly function as an LOCC monotone.  
To address this, we clarify an important distinction: the counterexample presented in \cite{Hayden:2023yij} involves a mixed-state density matrix \(\rho_{\textrm{\tiny{ABC}}}\). In contrast, for a pure tripartite state \(\ket{\psi}_{\textrm{\tiny{ABC}}}\), demonstrating monotonicity under a partial trace is straightforward.  
To see this more explicitly, we turn to the subadditivity relations for \(A, A^*\) and \(B, B^*\), which serve as the
\begin{align}
    I(A:A^*)\geq0 \implies 2 S(A)\geq S_R(A:B)\\
    I(B:B^*) \geq 0 \implies 2 S(B) \geq S_R(A:B)
\end{align}
Consider a pure state $\ket{\psi}_{ABC}$ then we note the following results
\begin{align}
    S_{R}(A:BC)=2 S(A)\\
    S_R(B:AC)=2 S(B)
\end{align}
The above results when combined with the upper bound for $S_R(A:B)$, it is easy to show that
\begin{align}
    S_R(A:BC)\geq S_R(A:B)\\
    S_R(B:AC) \geq S_R(A:B)
\end{align}
which is the criteria to be met for \(S_R(A:B)\) to serve as a valid measure of correlations.  Thus, it is clear that as long as \(ABC\) is in a pure state, the reflected entropy remains monotonic under a partial trace. Since our discussion in this article is limited to pure states, this ensures that no inconsistencies arise.

\section{Generalized L-entropy}\label{gen_L}
In this section, we define the generalized latent entropy, $L_{\text{gen}}$, as a hierarchical refinement of the $L$-entropy presented in \cite{newlentropy} mentioned in the main article. For a given $n$-party pure state, the generalized L-entropy can be constructed utilizing the following steps.

\begin{enumerate}
	\item For an $n$-party pure state $\rho_{A_1\cdots A_n}$, construct all reduced density matrices $\rho_k$ with $2\leq k\leq \text{Max}\left[2,\lfloor n/2 \rfloor\right]$. 
	\item Canonically purify each $\rho_k$ as $\rho_{A_1\cdots A_k} \;\longrightarrow\; \ket{\sqrt{\rho_{A_1\cdots A_k}}}_{A_1\cdots A_k A_1^*\cdots A_k^*}$.
	\item Compute the reflected entropy $S_R(\mathcal{A}:\mathcal{B})$ by bipartitioning the $k$ parties into two complementary subsets: $\mathcal{A}=A_1A_2\cdot\cdot\cdot A_{\lfloor \frac{k}{2}\rfloor}$ containing $\lfloor k/2\rfloor$ parties and $\mathcal{B}=A_{\lfloor \frac{k}{2}\rfloor+1}A_{\lfloor \frac{k}{2}\rfloor+2}\cdot\cdot\cdot A_{k}$ containing the rest. It is possible to choose $\mathcal{A}$ and $\mathcal{B}$ in $n_k$ different ways where $n_k$ is given as,
	$$
	n_k =
	\begin{cases}
		\dfrac{1}{2}\binom{k}{k/2}, & k \ \text{even}, \\[6pt]
		\binom{k}{\lfloor k/2\rfloor}, & k \ \text{odd}.
	\end{cases}
	$$
	\item Utilizing the upper bound of the reflected entropy, define $\ell_k^{(i)}$ as,
	\begin{align}\label{genell}
		\ell_k^{(i)} = 2 \min\{S(\mathcal{A}), S(\mathcal{B})\} - S_R(\mathcal{A}:\mathcal{B}),
	\end{align}
	where $i$ denotes a particular combination of $\mathcal{A}$ and $\mathcal{B}$ among $n_k$ different ways.
	\item For fixed $k$, consider the geometric mean of all $\ell_k^{(i)}$s
	\[
	L_k = \Big(\prod_{i=1}^{n_k} \ell_k^{(i)}\Big)^{1/n_k}.
	\]
	\item Finally, the generalized latent entropy is defined as
	\[
	L_{\text{gen}}(A_1\cdots A_n) = \Big(\prod_{k=2}^{\lfloor n/2 \rfloor} L_k\Big)^{1/\lfloor n/2 \rfloor}.
	\]
\end{enumerate}
Note that, this generalized construction can also be used for $n\leq5$ party systems which exactly results to the formulation explained in main article. It can be easily shown that the generalized L-entropy satisfies all the conditions to qualify as a GME measure. Interestingly, the generalized L-entropy provides a hierarchy of $k$-uniform states according to amount of GME present \cite{newlentropy}. In this context, the $\lfloor\frac{n}{2}\rfloor$-uniform states, known as absolutely maximally entangled (AME) states, yield the presence of maximum possible GME in $n$-party pure states.

	\section{L-entropy in CFT}\label{L_cft}
Consider a 2D CFT on a circle of circumference $L$, partitioned into three contiguous arcs $A$, $B$, and $C$ arranged sequentially. 
Let $l_A$ and $l_B$ denote the lengths of $A$ and $B$, respectively.
The bipartite L-entropy for the adjacent pair $(A,B)$ is
\begin{align}
	\ell_{AB}^{(\mathrm{circle})}
	=\frac{c}{3}\min\!\bigg\{
	\log\!\Big[\frac{L}{\pi\epsilon}\frac{\sin\!\big(\pi\frac{l_A+l_B}{L}\big)\,\sin\!\big(\pi\frac{l_A}{L}\big)\,}
	{\sin\!\big(\pi\frac{l_B}{L}\big)\,}\Big],\,
	\log\!\Big[\frac{L}{\pi\epsilon}\frac{\sin\!\big(\pi\frac{l_A+l_B}{L}\big)\,\sin\!\big(\pi\frac{l_B}{L}\big)\,}
	{\sin\!\big(\pi\frac{l_A}{L}\big)\,}\Big]
	\bigg\}
	+ \mathrm{const}.
	\label{eq:Lcircle-min}
\end{align}
Here $\epsilon$ is the UV cutoff and the additive constant is non–universal. The overall prefactor $c/3$ encodes the universal logarithmic scaling governed by the central charge, as in single-interval entanglement entropy.

Furthermore, we consider the whole CFT difined on an infinite line by taking the limit $L\to\infty$ which yields $\sin\left(\frac{\pi x}{L}\right)\to\frac{\pi x}{L}$. In this limit the bipartite L-entropy can be computed as
\begin{align}\label{eq:Lline-min}
	\ell_{AB}^{(\mathbb{R})}
	=\frac{c}{3}\min\!\bigg\{
	\log\!\Big[\frac{(l_A+l_B)\,l_A}{l_B\,\epsilon}\Big],\,
	\log\!\Big[\frac{(l_A+l_B)\,l_B}{l_A\,\epsilon}\Big]
	\bigg\}
	+ \mathrm{const}'
\end{align}
 and defining  $l_{min}\!=\!\min(l_A,l_B)$ and $l_{max}\!=\!\max(l_A,l_B)$ in compact form,
\begin{align}
	\ell_{AB}^{(\mathbb{R})}
	=\frac{c}{3}\,\log\!\Big[\frac{(l_A+l_B)\,l_{min}}{l_{max}\,\epsilon}\Big]
	+ \mathrm{const}'.
	\label{eq:Lline-compact}
\end{align}
The constants in \eqref{eq:Lcircle-min} and \eqref{eq:Lline-compact} differ by scheme-dependent terms  and are non–universal, whereas the length dependence and the coefficient $c/3$ are universal for adjacent intervals.  Note that, in both these cases, the multipartite L-entropy will be computed by considering all the other bipartite L-entropy.

\twocolumngrid

\end{document}